\def\year{2022}\relax
\documentclass[letterpaper]{article} 
\usepackage{aaai22}  
\usepackage{times}  
\usepackage{helvet}  
\usepackage{courier}  
\usepackage[hyphens]{url}  
\usepackage{graphicx} 
\usepackage{natbib}  
\usepackage{caption} 
\DeclareCaptionStyle{ruled}{labelfont=normalfont,labelsep=colon,strut=off} 
\usepackage{algorithm}
\usepackage{algorithmic}

\usepackage{newfloat}
\usepackage{listings}
\floatstyle{ruled}
\newfloat{listing}{tb}{lst}{}
\floatname{listing}{Listing}
\title{Some Observations on Fact-Checking Work with Implications for Computational Support}
\author{Rob Procter\textsuperscript{\rm 1, \rm 5},
Miguel Arana-Catania\textsuperscript{\rm 2},
Yulan He\textsuperscript{\rm 3, \rm 5},
Maria Liakata\textsuperscript{\rm 4, \rm 5},
Arkaitz Zubiaga\textsuperscript{\rm 4},
Elena Kochkina\textsuperscript{\rm 4},
Runcong Zhao\textsuperscript{\rm 3}
}
\affiliations {
   
\textsuperscript{\rm 1}University of Warwick\\
\textsuperscript{\rm 2}Cranfield University\\
\textsuperscript{\rm 3}Kings College London\\
\textsuperscript{\rm 4}Queen Mary University of London\\
\textsuperscript{\rm 5}Alan Turing Institute for Data Science and AI\\
rob.procter@warwick.ac.uk, miguel.aranacatania@cranfield.ac.uk, yulan.he@kcl.ac.uk, m.liakata@qmul.ac.uk, a.zubiaga@qmul.ac.uk, e.kochkina@qmul.ac.uk, runcong.zhao@kcl.ac.uk
}

\begin{document}

\maketitle

\begin{abstract}
Social media and user-generated content (UGC) have become increasingly important features of journalistic work in a number of different ways. However, the growth of misinformation means that news organisations have had devote more and more resources to determining its veracity and to publishing corrections if it is found to be misleading. In this work, we present the results of interviews with eight members of fact-checking teams from two organisations. Team members described their fact-checking processes and the challenges they currently face in completing a fact-check in a robust and timely way. The former reveals, \textit{inter alia}, significant differences in fact-checking practices and the role played by collaboration between team members. We conclude with a discussion of the implications for the development and application of computational tools, including where computational tool support is currently lacking and the importance of being able to accommodate different fact-checking practices. 
\end{abstract}

\section{Introduction}
The study reported in this paper was conducted in the context of the PANACEA project
\emph{PANdemic Ai Claim vEracity Assessment: An AI-enabled evidence-driven framework for claim veracity assessment during pandemics}\footnote{\url{https://panacea2020.github.io/}}. 
The main objective of this project was to analyse the pandemic from an informational and communicative perspective, and to develop analysis and verification tools to combat the misinformation crisis. Its outcomes included new fake news datasets \cite{arana2022natural, dougrez2022phemeplus}, evaluation of existing and development of novel fact-checking techniques \cite{kochkina2023evaluating, arana2022natural, dougrez-lewis-etal-2021-learning, zhu2022disentangled, abumansour2021qmul, zhang2021supervised, si2021topic, wenjia2023} and a prototype fact-checking tool \cite{zhao2023panacea}.

In this paper we present some findings from interviews with fact-checkers about how they currently tackle the problem of misinformation, the methods they follow, tools they currently use and how they think NLP-based techniques and tools might assist them to do this more effectively given the pressures they are under and the human resources at their disposal.

\section{Related work}
Social media and user-generated content (UGC) are increasingly important features of journalistic work in a number of different ways \cite{Tolmie2017supporting}. For example, they are used as primary sources for leads and for the discovery of stories \cite{brandtzaeg2016emerging}. However, the growth of misinformation means that news organisations have had to devote more and more resources to determining the veracity of such content and to publishing responses if it is found to be misleading \cite{Bontchevasession}.

The growing challenges of dealing in a timely way with the escalating volume of misinformation have resulted in the development of a range of different NLP-based techniques and tools to support fact-checking work \cite{lazarski2021using, zeng2021automated, guo2022survey,10.1162/tacl_a_00503}. However, while several reviews of the fact-checking research literature have been published (e.g., \citet{demartini2020human}, \citet{das2023state}), with a few notable exceptions (e.g., \citet{micallef2022true}, \citet{juneja2022human}), there has been relatively little primary research aimed at understanding in detail the daily challenges fact-checkers face when dealing with, for example, COVID misinformation or 'infodemic' \cite{diseases2020covid}. \citet{belair2023knowledge} interviewed fact-checkers and major social media platform workers to explore how they collaborate on fact-checking. \citet{micallef2022true} interviewed a total of 21 fact-checkers about their methodologies, use of tools and challenges. They found that current fact-checking practices are largely manual, ad-hoc, and limited in scale, scope, and reach.


In earlier work, we conducted detailed ethnographic studies of of journalists working in newsrooms with a particular focus on their use of UGC when looking for leads to candidate stories for publication \cite{Tolmie2017supporting}. One of the key findings of these studies was that fact-checking is not a ‘one stop shop’ but a concern that becomes relevant again and again throughout the process of researching and writing a story. We concluded that tools need to support fact-checking as ongoing work, including making visible what fact-checking work has already been undertaken and by whom. 

At the time this study was performed, the newsroom we studied had yet to establish a dedicated fact-checking team, relying instead on journalists to integrate this within their normal practice. With the growth in scale of misinformation years, however, fact-checking has emerged as a distinctive practice in its own right \cite{graves2016rise}. Hence one of the objectives of the PANACEA project was to understand how members of fact-checking teams go about doing this work, the challenges they face and what role(s) fact-checking support tools might then take within the overall fact-checking workflow \cite{juneja2022human}.  

\section{Methodology}
A key element of the project was working with fact-checking experts to understand in depth the challenges of dealing with misinformation, as well as its unique characteristics in the case of the pandemic. Our aim was to carry out this analysis as comprehensively as possible, and therefore the project involved partners with very different expertise and approaches to responding to the challenges of misinformation at scale. These included a UK mainstream news organisation and Full Fact\footnote{\url{fullfact.org/}}, an independent, UK fact-checking organisation.


Originally, our methodology was to have been based on a new series of ethnographic studies of fact-checking work, with the aim of observing and understanding in detail the ways in which the work is accomplished, moment by moment \cite{rouncefield2016ethnomethodology}. However, the COVID pandemic made that impossible, so, instead, we conducted a series of semi-structured interviews with members of fact-checking teams within the UK mainstream news organisation and Full Fact. A total of eight interviews were conducted in late 2021. Interviewees included four fact-checkers, who were also trained journalists, two news editors and two data scientists. The aim was to identify key stages in the fact-checking process, the activities and people involved in each stage, and the key challenges they currently face. Finally, we invited interviewees to suggest how AI tools might help to meet those challenges and how. An outline of the interview questions can be found in the Appendix. 

Ethics approval for the study was obtained through the Warwick University Biomedical and Scientific Research Ethics Committee.

The interviews lasted for approximately 60 minutes on average. They were recorded and transcribed and then analysed by the interviewer. The top level themes were defined \textit{a priori} by the interview questions based on earlier research (\cite{Tolmie2017supporting}) and literature reviews. Interviewee responses where then analysed inductively  using Grounded Theory \cite{birks2013grounded} to explore these themes in depth.
In the next section, we report on the results of these discussions with fact-checkers.

\section{Findings}
In this section we report on the results of the analysis of the interviews. The findings are structured around the principal stages of the fact-checking workflow and are illustrated by sample extracts from the transcripts. Interviewees and names of individuals and/or accounts mentioned by interviewees have been anonymised.

\subsection{Surfacing claims}
Before fact-checking can begin, candidate i.e., 'check-worthy' claims need to be identified. Searches of social media are often a key part of this, and the interviewees explained how this is typically done. 

\begin{quote}
``OK, let’s talk Twitter. So we’ll identify the concept we’re looking at or sometimes from a trend that we just notice and there’s various ways we can go about it. The trending topics themselves might be useful. Just the Twitter search. Then zip back on a timeline from there, from their results, to see if it’s mentioned particular words and then drill down to individual posts \ldots most times you can find the first posting of a viral claim. Sometimes it’s somebody who got virtually no traction on it. Someone else picked upon It and made their own. Various people could be thinking the same thing at the same time, so it’s difficult to see who the first people were sometimes.''
\end{quote}

During an event such as the COVID pandemic, fact-checkers become aware that fake news articles are likely to appear on that topic, so their attention becomes more focused.

\begin{quote}
``Yeah, we do keep a very specific eye on certain channels on social media just to see who’s promoting what. Like you are seeing couple stories over a few weeks about these fake NHS vaccine letters being sent to schools\ldots So, we are keeping an eye on how viral they were getting on social media\ldots there’s been a couple of stories on vaccine letters just to reassure parents and pupils. And then just keeping an eye on those demonstrations against it.''
\end{quote}

As an event such as the COVID pandemic evolves over time, fact-checkers are likely to become aware of specific accounts that are particularly active in spreading disinformation and these accounts will then be the target of more focused monitoring.

\subsection{Selecting which claims to check}
Once candidate claims have been identified, decisions then have to be made about which ones are to be prioritised for fact-checking. In both organisations, these decisions are made initially at the start of the daily news round. Importantly, these decisions are the result of collaborative effort, relying on the collective expertise of editors and fact-checkers, as a data scientist at Full Fact observed.

\begin{quote}
``In the morning fact-checkers will choose what to check by discussion amongst themselves. It does vary a lot. Every claim is very different, but often will include things like working out exactly who said what, what exactly is the claim, what’s the audience, what’s the context?''
\end{quote}

In the following extract, the editor at Full Fact explained how this is done.

\begin{quote}
``The team will discuss the various claims that are put forward. We will look at things like how widely have they been shared? Who made the claim? How prominent is that person? Is it as a member of the public, is it a politician? Was it the front page of a national newspaper? Or was it page 36. And also, we’ll look at basically how high a harm claim was it, so, is it you should inject bleach to avoid getting coronavirus, in which case potentially very high harm. If it is inaccurate and based on all of those various different factors will prioritise the pieces that we can work on any given day\ldots''
\end{quote}

A similar process is followed by the mainstream news organisation, as an editor explained. One significant difference, however, is that the fact-checking team can rely on support from the news organisation's journalists, so monitoring for claims to check continues during the course of the day. 

\begin{quote}
``We’re all kind of will constantly jabbering to each other on various different chat apps, but we also have two editorial meetings a day\ldots
people from all of our bureaus who will come together \ldots and they’ll say this is happening, this is one of the thematic teams I run. I run another thematic team and it means that we can pull out those threads and say, OK, that’s interesting\ldots sometimes when you get counter narratives as well, we can kind of join those dots together. That’s kind of how it works from a monitoring perspective.''
\end{quote}

Essentially, the process the two editors describe is that of triaging. They remarked that, given the limited resources available to them, the amount of effort required will place a limit on the number of claims chosen for fact-checking each day. One news editor outlined three claim dimensions used to assess whether a claim makes the check-worthy list: \textit{spread}, \textit{severity} and \textit{amplification}.

\begin{quote}
``The three tests for me are spread, severity and amplification, and we’ve all kind of come to the same conclusions across the disinformation world. Spread is not just how widely is this being spread. It’s also things like has it crossed into different languages and is it moving around the world\ldots are people engaging with it and severity, is this something that actually is going to cause either real-world harm or did it come from somebody who could have a significant impact on people? The obvious example is Donald Trump and his injecting bleach, but another one which is quite useful is Novak Djokovic talking about anti-vaccine and anti-COVID stuff \ldots something like that is significant because he’s not the sort of chap you think is going to say something [like that]\ldots and he had access to an audience and to followers who weren’t probably part of the conspiracy world. So, when something gets amplified by a celebrity, particularly a celebrity like him who’s considered to be sort of straight dealing, healthy, a good role model. Then for me that ticks the severity box but also the whole amplification thing\ldots.''
\end{quote}

On the question of \textit{spread}, a fact-checker in the mainstream news organisation commented on some of the deficiencies of the tools they currently use for searching on social media. 

\begin{quote}
``I mean we're really looking for engagements. Usually, if something is being tweeted around a lot but it’s not being engaged with, it’s probably not much of a discussion, but if something is getting a lot of engagement, lots of retweets, lots of comments, that is worth picking up on. And so, we have to look at it. I mean, there’s all sorts of things which search tools can’t really handle, and one of them is satire and sarcasm. So, if you’re trying to do something that’s looking for sentiment analysis it often misses this completely because they don’t understand the British way of repeating back something sarcastically. And that has been a problem for us in the past.''
\end{quote}

The issue of \textit{severity} rests on trying to estimate the likely `real-world consequences' or harm, as a news editor explained.

\begin{quote}
`\ldots seeing when things are starting to go from being a load of nonsense. A few people are talking about to something actually that could have real world consequences, and that’s when it comes up in our daily meeting that we have and someone like me makes the call as to whether we pursue it or not.''
\end{quote}

Claims that are not selected for fact-checking may continue to be monitored and come back under consideration at a later date, as a fact-checker at the mainstream news organisation explained.

\begin{quote}
''One to be published today on ivermectin has been in the works for months. I mean, yeah, so that probably didn’t cross the threshold for quite a while until large interest groups became interested and then it turned into a multi million dollar trade for anti-vaccine groups who actually made money from it so that crossed the threshold when people started rushing out to buy it. So yeah, there the potential harm threshold basically was reached.''
\end{quote}

Fact-checkers gave examples of a candidate claim where the decision to fact-check rested on a quite nuanced application of the \textit{calculus of fact-checking worthiness}. This turned out to be a common situation fact-checkers have to deal with. In this first example, for the fact-checker the issue concerns errors in the evidence quoted in support of the claim rather than the truth of the claim itself, along with deliberation on the balance between likely spread of the claim and the extent of harm if not corrected. Note also how the fact-checker drew on a previous fact-check in resolving this new claim.

\begin{quote}
``The other one I wrote this week is about [\ldots] and her TV show, on which she and [\ldots] claimed that 90\% of people in hospital with COVID are un-vaccinated. That’s a good example of something I already knew wasn’t true because I’ve written about this subject fairly recently. The real number is about 36\%, so you could certainly say that number is substantially misleading. What they were basically saying is you should get vaccinated because it seriously reduces your chance to go into hospital and then they use this 90\% numbers as substantiation for that. So, their basic point is right\ldots you should get vaccinated because it really does substantially reduce the chance of going to hospital but the number they used to substantiate it was wrong. That would be an example of something that was important to check because it was on national TV. It’s been made numerous times, but the potential for harm is probably lower.''
\end{quote}

In this next interview extract, a fact-checker provides an example of a claim that requires some deliberation about what is being claimed in order to determine whether it meets the threshold for fact-checking and where the decision hinges again on weighing up the spread/harm balance.

\begin{quote}
XXX [a well known author] wrote an article in The YYYY [a weekly UK magazine] in which [XXX] said that the vaccines can’t stop Corona virus spreading, and you could sort of interpret that in two different ways, you could say: ‘Well, not every single person who gets vaccinated is perfectly protected from catching or spreading it so vaccination doesn’t completely stop all examples’, but maybe that isn’t what she meant. Maybe like ‘it can stop it because there are loads of times in an individual case when it does stop it from spreading, someone doesn’t catch it in the first place’. So which is the meaning of that sentence that we would be checking? In the context of that article, I think it was clear she was saying the second one, she was saying that it doesn’t make any difference at all to reducing the spread, and that is untrue because we can prove that’s untrue, whereas the first possible meaning would be true\ldots So, we have a lot of detailed conversations like that \ldots it’s probably not going to be seen by as many people\ldots, but the potential for harm is much higher because [XXX] was carefully constructing an argument based on data where she explained that being vaccinated didn’t make any difference on spread, and if people believe that and take it seriously it could be really harmful.''
\end{quote}

Finally, \textit{amplification} is the third factor in fact-checking teams' deliberations about whether to proceed with a specific fact-check. As a news editor explained, this relies on the team being able to estimate risk that publishing a fact-check may raise the profile and thus increase public awareness of a claim, which might then be counter-productive to the impact of the fact-check itself.

\begin{quote}
''\ldots the whole amplification thing is something that we wrestle with all the time and because you know, because of having the [mainstream news organisation's] platform, you know if we say, oh this is happening and we have to be really careful about it, particularly with [XXX] and other conspiracy theories \ldots We want to report it without making it seem appealing.''
\end{quote}

It is clear from this that amplification is a factor that will be revisited when the fact-check is written up for publication.

\subsection{Finding the evidence}
Once decisions have been made about which claims meet the check-worthiness threshold, the actual fact-checking work then begins. As we will see in the extracts below, the interviews revealed a significant diversity in the methods used. At Full Fact, the approach taken generally relies on assembling relevant documentary sources, which are then reviewed to see whether the evidence gathered is consistent with the a claim. This process begins with searching for sources online. As a fact-checker in fact-checking organisation explained, this is where the effort escalates.

\begin{quote}
``Sometimes, even if you know what sort of broad topic area, there are still so many sources of information on that topic to sift through. Something that would simplify that process, or at least pull out a list of what you might want to look at would definitely make that process easier. And similarly, the experts to talk to, here’s some people you might want to consider, or some organisations, that kind of thing\ldots Primarily the first thing that we’ll check for is have they been fact-checked previously? Has this person shared misinformation? Is that something that we need to aware of? And what sort of expertise that they have in an area, that kind of thing.''
\end{quote}

As the extract above reveals, scientific papers or reports may not always be judged to be sufficient in and of themselves to determine the truth of a claim. Sometimes, a fact-checker will find it necessary to talk to an acknowledged expert in the field.

\begin{quote}
``In a lot of cases it may involve reading published reports\ldots In more technical cases it may also have involved speaking to an expert. A lot of these pandemic related stories, they have spoken to a doctor or medical researcher ‘how do you know MRNA vaccines are actually safe?’\ldots Get expert opinion, or explanation. ''
\end{quote}

As the extract below illustrates, the decision on what approach to take for fact-checking a claim involves weighing up a number of factors, including familiarity with the topic.

\begin{quote}
``We have expert voices in our pieces to different extents, some pieces it may just be a straight write up of someone made a claim about a statistic about poverty, for example. And here’s what these statistics about poverty say. Here’s any caveats you may need to know about this data\ldots We don’t necessarily need to go and talk to an expert about it. Often, we’ll speak to experts if it’s a broader claim\ldots If it’s less specific, or if it’s a topic that we’re less familiar. You do start to build up topic expertise as you go through fact-checking. If it’s something that we’re less familiar, we often go to experts even just to say 'where should we be looking for this information? And this does this sound right?'. Once we’ve spoken to experts, whether they’ve pointed us towards information, or they’ve given their view on the topic, we will present that and we’ve already tried to assess how authoritative that person is in that field, how much that matches up with the consensus view on that topic, if there is one.''
\end{quote}

In contrast, in the mainstream news organisation, for organisational reasons, talking to experts was the default fact-checking approach as this was seen to be the way to preserve their reputation for impartiality. 

\begin{quote}
``We would always seek to speak to an expert\ldots you can get the data if you wanted to find out if somebody said this number of people have died from COVID in in the US. You could go to the Johns Hopkins stats or whatever. But if we were doing something that was in any way more detailed what we can’t say this is coming from our mouth\ldots impartiality is very important\ldots we would always look to speak to somebody who is an expert\ldots [then we can say] we've spoken to X expert or Y expert and this is what they said.’\ldots Always seek to speak to an expert and multiple experts if possible. Seldom would regard one source as being sufficiently credible to be relied on alone.''
\end{quote}

When searching for credible experts, fact-checkers have to be wary of the lengths some sources of disinformation will go to disguise their true stance on issues of the day.

\begin{quote}
``There’s lots of groups. They set themselves up as sounding as though they are pro climate change and they are anti climate change so a journalist doing something at pace looks at the Coalition for climate confidence or something like that. And you think oh, that’s interesting\ldots [but] it transpires that actually it’s a kind of coalition of coal producers\ldots''
\end{quote}

Interviewees emphasised how fast paced the process of fact-checking had to be and the pressure to produce a result can lead to cutting corners in ways that could make the process less thorough.

\begin{quote}
``You get that concern about sort of built-in subconscious bias If you keep going back to the same sources the whole time. Because as a journalist you’re always doing things that 1000 miles an hour, and you know a certain website is easier to read. It’s kind of more likely to confirm what it is that you’re wanting to do or whatever, and so actually you get slightly lazy and don’t then kind of really start checking things out more broadly.''
\end{quote}

In some cases, finding which experts to talk to is quite straightforward. fact-checkers retain a knowledge of who to talk to from previous fact-checking exercises as this news editor explained.

\begin{quote}
``I mean again now because we’ve been doing COVID for so long. We have people that we know that we can go to and we know that they’re trustworthy.''
\end{quote}

When a new claim breaks, however, then the challenge is to find people who measure up to the news organisation's standards for being an expert. A fact-checker in the mainstream news organisation described an example of such a fact-check.

\begin{quote}
``When there’s something comes up that actually we’re not that knowledgeable at about. You know, we’re all researchers, we’re all journalists by trade, we know how to do this sort of stuff, but particularly when something comes up that isn’t necessarily your area of expertise. When the 5G stuff started, you do find yourself Googling, trying to work out who the kind of national group is for 5G operators and that sort of thing, and it might be that they’re based in Berlin,\ldots and so you are spending a lot of time trying to find the right people.''
\end{quote}

The Full Fact editor described in more detail the ways in which they go about identifying credible experts, revealing how lengthy this process can be.

\begin{quote}
``A lot of the things that we do end up fact-checking have gotten quite a bit of coverage either in the news or by other fact-checkers\ldots people who have spoken to other fact-checkers. That at least gives you a sense of whether or not the person is willing or has time to speak to media organisations. Then you have then got to go and assess whether or not we think that they are a good source of information. That often can involve some digging into their qualifications. Are they an academic, what allows them to speak with authority on this topic and would make us happy to say that they are an expert in that topic area\ldots We have a look to see what their professional background was as far as that was available. We’d look and see what they had previously said on the topic a) because they may have spoken on the thing that we’re looking for exactly already, but also we’re very conscious that we want to try and make sure that we’re talking to independent organisations. And be aware if the experts that we’re putting forward have skin in the game and so it’s trying to identify anything that might be a red flag.''
\end{quote}

One way to make this process quicker is to build up a network of relationships with people and organisations that have a reputation for being credible and trustworthy.

\begin{quote}
``We may speak to organisations that we have in our network. And if they’re not suitable, say, OK, well, we’re looking for someone on this topic. Is there anyone that you can think of that might be useful there for this particular fact-check. And then if all that fails, there’s just a little bit of again having to do the research trying to identify people just looking at, for example, academic departments, that kind of thing. And then just cold calling people, talking to a University Press Officer saying, is there anyone that you can put forward to speak on this topic or think tanks that work in the specific fields that we might be researching.''
\end{quote}

As fact-checkers accumulate experience of the kinds of topics that feature in disinformation campaigns, they grow their networks of trusted experts, making finding one suitable for a specific fact-check more straightforward.

\begin{quote}
``It’s something that we do encounter less now as we’ve grown as an organisation that we have managed to build up larger networks ourselves. But definitely when we have encountered that situation where we’re starting from scratch, it can be quite time consuming.''
\end{quote}

Previous fact-checks on a topic can also be useful when a new claim appears on the same or related topic, especially when the topic is unfamiliar to the fact-checker who has been allocated it.

\begin{quote}
''If we can’t establish that something we’re reading online is authoritative as a source for information, then it can be hard to use. One very common source that we would all use individually for information on the subject that’s unfamiliar to us is a previous fact-check, one of our own\ldots once [a claim's] been through this very laborious editing process and actually got published, it’s pretty reliable, and if someone has previously written an article on the same subject with all the links that you might need taking you to the places where the documents can be found, that’s a brilliant way of getting to be familiar with the subject.''
\end{quote}

Evidently, apart from the benefit of expediting a new fact-check, records of previous fact-checks serve as a way of sharing knowledge and building expertise within the team.

\subsection{Reviewing the fact-check}
The results of a fact-check are then reviewed by members of the team to make sure that no mistakes have been made, as a fact-checker in Full Fact explained. 

\begin{quote}
``The reviewer virtually does the same amount of work as the writer because they’re repeating the entire process of building up the evidence base\ldots which can be a very, very complex, sometimes drawn out process\ldots Collaboration is really, really important. Mainly to check the writer has got it all right but also it allows more than one person to understand the subject.''
\end{quote}

The importance of reviewing fact-check results was underlined by the Full Fact data scientist.

\begin{quote}
``Each article has at least one person who reviews it, and that basically is going through all with a fine-tooth comb. Looking at, you know if there’s any data or analysis, are we happy that we’ve interpreted all that correctly. Have we done the calculations correctly? Have we linked to all of our sources of information, is our grammar right\ldots The editing process, the reviewing stage, then it’s very collaborative.''
\end{quote}

The Full Fact editor expanded on this.

\begin{quote}
``That’s sort of our standard, there has to be 3 pairs of eyes on it. And if it’s a more complicated piece, if it’s a piece that has a lot of follow up work or media work involved in it that number can escalate. It could be at times anything up to even eight or ten people if it’s a particularly complex piece that we want to do a lot of work around. But at minimum we say three people now have to be involved in fact-checking\ldots Depending on how sensitive a piece is will depend on how senior a person in the team has to be that third pair of eyes.''
\end{quote}

For the mainstream news organisation, the fact-check review focuses on establishing that due diligence has been done in the selection of expert opinions. 

\begin{quote}
''We have people that we know that we can go to and we know that they’re trustworthy and all the rest of it. And but yeah, you know we would always check, check and double check, you just can’t get this stuff wrong because if you’re saying that other people are peddling disinformation and then you get something wrong. It’s an open goal \ldots If [a fact-checker] came to me and said 'here’s a piece that I’ve written' I would say, well, who did you speak to? \ldots it would pretty much be my first question.''
\end{quote}

Clearly, for both organisations, failure to conduct a thorough review could have very damaging consequences for their reputation for trustworthiness.

\subsection{Writing up the fact-check}
The final stage is writing up the outcome of the fact-check for publishing. Here, a number of organisational guidelines come into play. Some kinds of claims are straightforward as in the example  given below by a data scientist in Full Fact.

\begin{quote}
``Basically, claims that are quantitative and which have a clear, a trustworthy ground source that we can verify against. This is just things like official government figures about macroeconomics. Things like the current rate of inflation. Three different measures of inflation are published every month. If someone says that retail price inflation is down 5\%, either that’s right or that’s wrong\ldots''
\end{quote}

The writing up of other kinds of claim may require a more nuanced approach. Both organisations emphasise that it is essential to present the findings in a way that respects the principle of balance, as the next four extracts emphasise. The first addresses balance, an issue about which news organisations are very conscious.

\begin{quote}
``I think one challenge for the public is how do you judge whether two sides of story are equal or balanced or biased. For example, there’s been complaints about how the [\ldots] presented both sides to debates that are really settled on. Things like climate change. You could have somebody who’s going to deny climate change, but do you give them equal airtime to somebody who is saying that climate change is real. And probably if you give that sort of false equivalence, false balance, the public comes away thinking, nobody really knows, but actually the evidence is most people say is overwhelming\ldots''
\end{quote}

The next three extracts stress that the presentation of the findings of a fact-check must allow the reader to make up their own minds. The Full Fact editor explained that it is not their goal to influence the reader.

\begin{quote}
``We aren’t trying to make anyone’s mind up on a topic within article, and the idea is to present the information to them and allow them to make up their own mind essentially about the claim and the topic and the claim and all that kind of thing. So, you know, other than to give a steer of this, this is the summary of what we’ve written about. We’re not trying to influence anyone.''
\end{quote}

Both fact-checking teams devote a considerable amount of effort to ensure an article's neutrality. In the following quote, a team member described how this is done at Full Fact.

\begin{quote}
``A lot of the work is writing up what’s been found. The writing of an article, which has to be very neutral, very clear, easy for anyone to read, and presenting all of the evidence in a kind of non-judgmental way so readers can then decide themselves whether not they believe this claim. And that process of writing and editing and publishing is quite collaborative. I think three fact-checkers will read a piece before it gets published to make sure that it is neutral, there’s no claims there that should be verified more carefully.''
\end{quote}

In this third extract, a Full Fact data scientist suggested that while AI can certainly assist in finding evidence for a fact-check, its presentation demands human skill and expertise.

\begin{quote}
``Our audience\ldots don’t want someone saying this is true, take our word for it. They want us to do the research for them\ldots then let make their own mind up. Can you get a machine to find evidence that a fact-checker can then filter and present it to the public? That’s a feasible goal. But automation of whether something is true or false depends so much on real-world understanding. It’s a very, very hard thing for any AI to do.''
\end{quote}

An editor in the mainstream news organisation explained that once the fact-check of a particular claim has been completed, the amplification test will be revisited before the claim is written up and will determine how it is written up if the decision is then made to publish it.

\begin{quote}
''\ldots Is the story in a public interest for some things are we amplifying behaviour or are we likely to encourage bad behaviour, so we have to put things to an editorial test before we report on them \ldots We want to report it without making it seem appealing. So, there are kind of various things that we do, you know we don’t live link. We overlay things that’s say false or misleading or whatever, and we’re really careful about not leaving too many digital breadcrumbs so that people can find you know the last thing we want is people to go Oh, how fascinating and kind of go down that rabbit hole. Yeah, so we have to be really responsible. And I think we’ve got a really keen sense of that.''
\end{quote}

In summary, the way that the results of a fact-check is written up is seen by both fact-checking teams as being just as critical to combating disinformation as ensuring that the fact-check is correct.

\subsection{Computational tool support}
Fact-checkers agreed that some kinds of claims could be checked automatically. However, they also were of the view that this would only be possible for a small fraction of the total they have to deal with. 

\begin{quote}
``There are certain types of claims which can be verified by machine\ldots claims that are quantitative and which have a clear, a trustworthy ground source that we can verify against. If someone says that retail price inflation is down 5\%, either that’s right or wrong, it’s the kind of claim which potentially can be checked by machine\ldots But even if it worked perfectly, it would only cover a tiny range of claims that are being made.''
\end{quote}

Hence, interviewees were keen to suggest other ways in which computational tools could support their work. A news editor in the mainstream news organisation explained how they could support surfacing candidate claims.

\begin{quote}
``In an ideal world something goes viral and we know it’s going viral because our marvellous bit of kit has said, oh, there’s a claim coming up that having taken the vaccine means that you’re more likely to have twins or something and access to a fact-checking tool, which could say, OK, here are six different sources and this is what they have to say on it.''
\end{quote}

The news editor went on to explain some of the benefits of this in more detail.

\begin{quote}
``for me getting a heads up about things that are becoming viral is the kind of golden goose for me, because we just have to wait until you know somebody tips us off to things. And missing things if they are so discrete to a certain community or a certain geographical area is a source of enormous frustration to me and I think also doesn’t allow us to be as diverse in our content as we would like to be, so something that could say, and I think [\ldots] are working on something around this, they’re starting to look at what they call claim like statements and how they bubble up. But that would be a real game changer for us\ldots''
\end{quote}

A fact-checker in the mainstream news organisation identified a deficiency in existing search tools.

\begin{quote}
''\ldots reliable sentiment analysis. Because the ones we’ve used in the past. OK, try hard bless them, but they seem to fall over on human emotion. And things like lying, those are the things that often frustrate the entire community that you know being able to go through lots of lots of data to find sentiment and to find if one is seriously engaged enough rather than that is a trivial thing.''
\end{quote}

The challenges of identifying claims that might be check-worthy has motivated the development by Full Fact of computational tools to assist in this process, as a data scientist in the fact-checking organisation explained, while stressing at the same time that automation of the whole fact-checking process is not a feasible goal.

\begin{quote}
``Speaking to fact-checkers it turns out that the decision process to decide what to check is very, very complex and relies lots on their expertise and intuition and so I’ve always refused basically to actually try to develop an algorithm to predict is this worth checking. Instead, what we try to do is develop tools which help surface claims in such a way that they can then decide is it worth checking it or not. So, what we do is we bring to attention of the fact-checker a variety of claims\ldots the reason that I never wrote a system to predict which claims were worth checking is because a fact-checker said, you know, this is our process. This is like hours of debate for every check. We can’t just automate that.'' 
\end{quote}

He went on to describe in more detail the approach taken to assist in identifying check-worthy claims, including what kinds of news stories are ignored and which ones are prioritised.

\begin{quote}
``One [tool] is essentially an index over the last 24 hours’ worth of UK media. So, take every sentence from a newspaper and say, is this making a claim?\ldots This algorithm basically will identify things like sports results, celebrity gossip, announcements of marriages and removes all of that from consideration. And it takes what’s left and roughly groups it by category, things like health, education, crime rates are broad categories. It’s just finding stories that are being talked about most often in newspapers. If a story’s picked up by several different journalists it will put that top of our list. The fact-checker is going to look at this this morning and say, are there any major stories in the broad area I’m interested in which I might have missed? And it takes few minutes to read it, which is important. Not demanding they spent hours learning this tool or using this tool is just five minutes going through the morning\ldots I think the last few months they have published like two or three fact-checks based on things found using that tool.''
\end{quote}

The data scientist then went on to describe how fact-checkers could be assisted in estimating the effort that would be entailed in a specific fact-check.

\begin{quote}
``Providing an accessible summary of the evidence around the claim, including sources and experts, trusted opinions or further sources of information in such a way that fact-checkers can very quickly decide this one is worth digging into a bit more. They still do the work themselves, which still requires this kind of expertise, you know, research and storytelling. But if you can give them enough evidence at the outset so that something that may look unpromising in itself, with a bit of data around it, may think, actually, that is worth looking into a bit more. That kind of tool, I think, is something that actually could have an impact.''
\end{quote}

A fact-checker in the mainstream news organisation suggested a way in which tools would be useful for situations where consulting their network fails to yield one or more suitable experts to talk to.

\begin{quote}
``You could get your algorithm to learn this has been used in articles 10 times in the last month as a credible source, that would be helpful because journalism’s the polar opposite of academia. We do things at pace and hope for the best\ldots Sometimes you’re going for tried and trusted sources so, anything that can give you a shortcut to doing would be hugely beneficial.''
\end{quote}

The Full Fact data scientist, using the example of finding evidence about a claim, reflected on the challenge of getting the balance right between the what tools can do and human expertise.

\begin{quote}
``I guess fact-checkers are already expert journalists that already engage in the field so they can be best helped by providing links to evidence faster than they would otherwise find it. Or maybe things otherwise they would have missed, but it’s quite challenging because basically it’s like a core part of their skill set and trying to automate that\ldots For example, I think a challenge for a tool is not just to find both sides of the story and present them but to actually somehow show where the balance of evidence is pointing, which fact-checkers are I think, skilled at and it’ll be interesting to see how you can automate that.''
\end{quote}

The Full Fact data scientist also acknowledged the concerns that introducing into the fact-checking workflow complex computational tools whose behaviour may be difficult for users to understand would itself raise issues of trust.

\begin{quote}
``There is the risk that the algorithms are a black box that nobody is going to really trust without a lot of experience. Explainability for fact-checkers would tend to be extracts, quotes and snippets and links to other sources. Primary sources. Things like links to the university homepage, saying yes, this person is employed by this university, here’s their homepage is kind of evidence. If a person has been talked about by other journalists, links and quotes from their articles and they’re quoted in the New York Times and in these 20 odd newspapers you know they’re widely spoken about in the media. You want to show the evidence, but you don’t want to reduce it down to a number or list, you want to say here is how they were quoted.''
\end{quote}

A news editor at Full Fact explained that their approach to improving fact-checking did not just involve the introduction of computational tools but making the process much more collaborative.

\begin{quote}
''One of the things that we’re trying to do over the next few years is not necessarily automate that process as such, but get a lot more people involved in the whole fact-checking process, get our supporters on board and more people involved. Basically open out from just being the members of the team who are involved in that process. We’re very conscious that there is a lot of misinformation out there. We are still relatively small team. We can’t check it all by ourselves. Bringing more people into that process, as well as looking for ways that we can improve it, automate it more. Exactly how we will do that’s still a work in progress, but that is that is our eventual aim.''
\end{quote}

Perhaps this emphasis on making the process more collaborative is not surprising given how teamwork already plays a key role in quality assuring the process.

\section{Discussion}
The interviews with fact-checking team members have revealed details of the different stages that collectively make up the fact-checking workflow and our findings broadly corroborate those of \citet{micallef2022true}. This begins with surfacing candidate claims, followed by reviewing these candidates in order to decide which ones are check-worthy. Here, an informal calculus based on \textit{spread}, \textit{severity} and \textit{amplification} comes into play in order to provide a degree of repeatability and quality assurance to the decision-making process, although the application of these factors is inevitably often quite nuanced. Hence, while tools have been developed to identify check-worthy claims automatically, it is evident that these cannot be a substitute for human expertise \cite{demartini2020human} and grasp of what it means to be in compliance with organisational policies \cite{bittner1965concept}. 

Importantly, the study reveals how the whole fact-checking process relies on collaboration. Not only is this important to ensure the process is robust but collaboration also supports the building of knowledge and expertise within the team. It is therefore not surprising that fact-checkers are skeptical about the feasibility of removing human expertise entirely from the process.

Following the identification of check-worthy claims, the actual fact-checking begins. Here, we observe variations in the process that are attributable to the different positions within the 'news and media ecosystem' that the two organisations occupy. This leads them to define their end goals in ways that may seem only very subtly different but which then 'back propagate' through the fact-checking workflow. 

The process followed by the fact-checking organisation generally focuses on searching for evidence to support or refute the claim in online sources such as scientific papers and documents published by credible and trustworthy organisations. 
In the case of the mainstream news organisation, the process relies on finding credible experts who are then willing to be quoted. This latter approach also has two advantages over the former. First, it enables the fact-checker to get answers that are tailored to the nature of the claim. Second, searches for evidence are likely to fail to resolve claims for which -- by their nature -- there is no record of them ever having been tested; talking to experts will then be the only option, especially when there is a compelling need to complete a check as quickly as possible. Some claims made during the COVID-19 pandemic, for example, (e.g., 'injecting bleach can cure COVID') may fit into this category. 

Regardless of methods, the goal of both organisations is to let the reader decide whether a claim is true or false, for Full Fact this means presenting a summary of credible evidence in a non-judgemental way, whereas for the mainstream news organisation, this means presenting the views of credible experts. 

Clearly, variations in fact-checking methodology need to be reflected in the design of tool support. With this in mind, we are currently developing tools to assist on the identification of experts that fact-checkers can then approach for an interview and a quote \cite{wenjia2023}. Irrespective, however, of the practices followed fact-checkers are very likely remember if they have done a check on a similar claim and this enables them to progress a new claim more quickly by consulting records kept either by themselves or other members of the fact-checking team. 

Once a claim has been resolved by the fact-checker, it is then reviewed by a second and sometimes even a third member of the team in order to make sure the process has been thorough, nothing has been overlooked and there are no errors, emphasising how essential team work is for quality assuring the fact-check. Once this has been completed, then the fact-check will be written up for publication. Neither organisation wants to be perceived as being the arbiters of whether a claim is true or false. Instead, the emphasis is on letting the audience decide and this generally means presenting the evidence gathered in a balanced way, not least because realising the value of of fact-checking depends on fact-checkers being trusted by their readers \cite{nakov2021automated}.

Interviewees stressed how challenging fact-checking can be, noting the growing volume of claims, the limited resources at their disposal and the pressure to complete the whole process as quickly as possible. This highlights that it would be beneficial for fact-checking tools to maintain a searchable repository of their results so that if a similar new claim appears on a similar topic, they can re-use a previous fact-check \cite{das2023state}.

In summary, fact-checking team members identified several ways in which tools might assist them in all stages of the fact-checking workflow. In summary, these include: surfacing claims, prioritising claims, searching for evidence and/or experts, and evaluating evidence and/or experts. In addition, one aspect of fact-checking work not mentioned in this respect but which would be interesting to examine, would be providing assistance in writing a balanced write up of the evidence for or against a particular claim. Finally, in our previous work, we suggested that it would be a mistake to provide support for fact-checking within a single tool or dashboard \cite{Tolmie2017supporting} and the diversity of practices we report here between dedicated fact-checking teams reinforces that view. 

\section{Conclusions and future work}
In this paper we have reported findings from a series of semi-structured interviews with news professionals working in fact-checking. We have seen how the fact-checking workflow consists of series of stages. Each stage has its particular challenges for fact-checkers working under pressure to determine quickly the veracity of a claim  and publish the result. It is also evident there is diversity in fact-checking practices and that robustness of the workflow depends on the collaborative and ongoing efforts of the team members, confirming findings from our earlier study \cite{Tolmie2017supporting}.

Interviewees identified ways in which computational tools could assist throughout the fact-checking workflow: surfacing claims, deciding on which are check-worthy and assembling the evidence. Interviewees were skeptical about the prospects for automating the process. They take for granted that the fact-checker is the \textit{human-in-the-loop} and that they must be the final arbiter when a claim is assessed. As with the introduction of advanced decision-making tools in other fields, there are important professional and organisational reasons why this should remain the case in fact-checking \cite{procter2022holding}. Hence, like several recent studies (e.g., \citet{demartini2020human}, \citet{rubin2022content} and \citet{das2023state}), we stress the importance of not attempting to automate fact-checking but to ensure tools are firmly based on an in-depth understanding of fact-checking work, including its collaborative dimensions. 

Any new tools must, of course, be easy to use and this also means that the challenges of explainability must be addressed if tools are to be trusted and thus accepted by professionals as fit for purpose \cite{das2023state, nakov2021automated}. These goals can be achieved through the use of ethnographic study methods \cite{rouncefield2016ethnomethodology} and the active participation of fact-checkers in design and development \cite{wolf2020ai}. The former is essential for an in-depth understanding of the work and its implications for the design of tools. The latter is essential to ensure that users' expectations of what computational tools are capable of are to be grounded in reality \cite{slota2020designing, arana2021machine}.

Finally, we argue that the way to address diversity in fact-checking methods \textit{and} the ongoing rapid advances in NLP techniques must be to focus on providing a \textit{fact-checking toolbox}, i.e., a set of inter-operable tools based on open standards that can be configured to meet the working practices of a particular fact-checking team and its members and can also be reconfigured as these practices change. 

\section{Acknowledgments}
We would like to thank the members of the fact-checking teams who generously gave of their time to support this project.

This work was supported by the UK Engineering and Physical Sciences Research Council (grant no. EP/V048597/1). In addition, YH and ML are each supported by a Turing AI Fellowship funded by the UK Research and Innovation (grant nos. EP/V020579/1, EP/V030302/1).

\bibliography{Mediate.bib}

\begin{thebibliography}{33}
\providecommand{\natexlab}[1]{#1}

\bibitem[{Abumansour and Zubiaga(2021)}]{abumansour2021qmul}
Abumansour, A.~S.; and Zubiaga, A. 2021.
\newblock QMUL-SDS at CheckThat! 2021: Enriching Pre-Trained Language Models
  for the Estimation of Check-Worthiness of Arabic Tweets.
\newblock In \emph{CLEF (Working Notes)}.

\bibitem[{Arana-Catania et~al.(2022)Arana-Catania, Kochkina, Zubiaga, Liakata,
  Procter, and He}]{arana2022natural}
Arana-Catania, M.; Kochkina, E.; Zubiaga, A.; Liakata, M.; Procter, R.; and He,
  Y. 2022.
\newblock Natural Language Inference with Self-Attention for Veracity
  Assessment of Pandemic Claims.
\newblock In \emph{Proceedings of the 2022 Conference of the North American
  Chapter of the Association for Computational Linguistics: Human Language
  Technologies}, 1496--1511.

\bibitem[{Arana-Catania, Van~Lier, and Procter(2021)}]{arana2021machine}
Arana-Catania, M.; Van~Lier, F.-A.; and Procter, R. 2021.
\newblock Machine Learning for Mediation in Armed Conflicts.
\newblock \emph{arXiv preprint arXiv:2108.11942}.

\bibitem[{B{\'e}lair-Gagnon et~al.(2023)B{\'e}lair-Gagnon, Larsen, Graves, and
  Westlund}]{belair2023knowledge}
B{\'e}lair-Gagnon, V.; Larsen, R.; Graves, L.; and Westlund, O. 2023.
\newblock Knowledge Work in Platform Fact-Checking Partnerships.
\newblock \emph{International Journal of Communication}, 17: 21.

\bibitem[{Birks et~al.(2013)Birks, Fernandez, Levina, and
  Nasirin}]{birks2013grounded}
Birks, D.~F.; Fernandez, W.; Levina, N.; and Nasirin, S. 2013.
\newblock Grounded theory method in information systems research: its nature,
  diversity and opportunities.
\newblock \emph{European Journal of Information Systems}, 22(1): 1--8.

\bibitem[{Bittner(1965)}]{bittner1965concept}
Bittner, E. 1965.
\newblock The concept of organization.
\newblock \emph{Social research}, 239--255.

\bibitem[{Bontcheva et~al.(2015)Bontcheva, Liakata, Procter, and
  Scharl}]{Bontchevasession}
Bontcheva, K.; Liakata, M.; Procter, R.; and Scharl, A. 2015.
\newblock Workshop on Rumors and Deception in Social Media: Detection,
  Tracking, and Visualization.
\newblock In \emph{Proceedings of the 24th International Conference on World
  Wide Web}.

\bibitem[{Brandtzaeg et~al.(2016)Brandtzaeg, L{\"u}ders, Spangenberg,
  Rath-Wiggins, and F{\o}lstad}]{brandtzaeg2016emerging}
Brandtzaeg, P.~B.; L{\"u}ders, M.; Spangenberg, J.; Rath-Wiggins, L.; and
  F{\o}lstad, A. 2016.
\newblock Emerging journalistic verification practices concerning social media.
\newblock \emph{Journalism practice}, 10(3): 323--342.

\bibitem[{Das et~al.(2023)Das, Liu, Kovatchev, and Lease}]{das2023state}
Das, A.; Liu, H.; Kovatchev, V.; and Lease, M. 2023.
\newblock The state of human-centered NLP technology for fact-checking.
\newblock \emph{Information Processing \& Management}, 60(2): 103219.

\bibitem[{Demartini, Mizzaro, and Spina(2020)}]{demartini2020human}
Demartini, G.; Mizzaro, S.; and Spina, D. 2020.
\newblock Human-in-the-loop Artificial Intelligence for Fighting Online
  Misinformation: Challenges and Opportunities.
\newblock \emph{IEEE Data Eng. Bull.}, 43(3): 65--74.

\bibitem[{Diseases(2020)}]{diseases2020covid}
Diseases, T. L.~I. 2020.
\newblock The COVID-19 infodemic.
\newblock \emph{The Lancet. Infectious Diseases}, 20(8): 875.

\bibitem[{Dougrez-Lewis et~al.(2022)Dougrez-Lewis, Kochkina, Arana-Catania,
  Liakata, and He}]{dougrez2022phemeplus}
Dougrez-Lewis, J.; Kochkina, E.; Arana-Catania, M.; Liakata, M.; and He, Y.
  2022.
\newblock PHEMEPlus: Enriching Social Media Rumour Verification with External
  Evidence.
\newblock In \emph{Proceedings of the Fifth Fact Extraction and VERification
  Workshop (FEVER)}, 49--58.

\bibitem[{Dougrez-Lewis et~al.(2021)Dougrez-Lewis, Liakata, Kochkina, and
  He}]{dougrez-lewis-etal-2021-learning}
Dougrez-Lewis, J.; Liakata, M.; Kochkina, E.; and He, Y. 2021.
\newblock Learning Disentangled Latent Topics for {T}witter Rumour Veracity
  Classification.
\newblock In \emph{Findings of the Association for Computational Linguistics:
  ACL-IJCNLP 2021}, 3902--3908. Online: Association for Computational
  Linguistics.

\bibitem[{Graves and Cherubini(2016)}]{graves2016rise}
Graves, L.; and Cherubini, F. 2016.
\newblock The rise of fact-checking sites in Europe.
\newblock Technical report, Reuters Institute for the Study of Journalism.

\bibitem[{Guo, Schlichtkrull, and Vlachos(2022)}]{guo2022survey}
Guo, Z.; Schlichtkrull, M.; and Vlachos, A. 2022.
\newblock A survey on automated fact-checking.
\newblock \emph{Transactions of the Association for Computational Linguistics},
  10: 178--206.

\bibitem[{Juneja and Mitra(2022)}]{juneja2022human}
Juneja, P.; and Mitra, T. 2022.
\newblock Human and technological infrastructures of fact-checking.
\newblock \emph{Proceedings of the ACM on Human-Computer Interaction},
  6(CSCW2): 1--36.

\bibitem[{Kochkina et~al.(2023)Kochkina, Hossain, Logan~IV, Arana-Catania,
  Procter, Zubiaga, Singh, He, and Liakata}]{kochkina2023evaluating}
Kochkina, E.; Hossain, T.; Logan~IV, R.~L.; Arana-Catania, M.; Procter, R.;
  Zubiaga, A.; Singh, S.; He, Y.; and Liakata, M. 2023.
\newblock Evaluating the generalisability of neural rumour verification models.
\newblock \emph{Information Processing \& Management}, 60(1): 103116.

\bibitem[{Krishna, Riedel, and Vlachos(2022)}]{10.1162/tacl_a_00503}
Krishna, A.; Riedel, S.; and Vlachos, A. 2022.
\newblock {ProoFVer: Natural Logic Theorem Proving for Fact Verification}.
\newblock \emph{TACL}, 10: 1013--1030.

\bibitem[{Lazarski, Al-Khassaweneh, and Howard(2021)}]{lazarski2021using}
Lazarski, E.; Al-Khassaweneh, M.; and Howard, C. 2021.
\newblock Using nlp for fact checking: A survey.
\newblock \emph{Designs}, 5(3): 42.

\bibitem[{Micallef et~al.(2022)Micallef, Armacost, Memon, and
  Patil}]{micallef2022true}
Micallef, N.; Armacost, V.; Memon, N.; and Patil, S. 2022.
\newblock True or false: Studying the work practices of professional
  fact-checkers.
\newblock \emph{Proceedings of the ACM on Human-Computer Interaction},
  6(CSCW1): 1--44.

\bibitem[{Nakov et~al.(2021)Nakov, Corney, Hasanain, Alam, Elsayed,
  Barr{\'o}n-Cede{\~n}o, Papotti, Shaar, and Martino}]{nakov2021automated}
Nakov, P.; Corney, D.; Hasanain, M.; Alam, F.; Elsayed, T.;
  Barr{\'o}n-Cede{\~n}o, A.; Papotti, P.; Shaar, S.; and Martino, G. D.~S.
  2021.
\newblock Automated fact-checking for assisting human fact-checkers.
\newblock \emph{arXiv preprint arXiv:2103.07769}.

\bibitem[{Procter, Tolmie, and Rouncefield(2022)}]{procter2022holding}
Procter, R.; Tolmie, P.; and Rouncefield, M. 2022.
\newblock Holding AI to Account: Challenges for the Delivery of Trustworthy AI
  in Healthcare.
\newblock \emph{ACM Transactions on Computer-Human Interaction}.

\bibitem[{Rouncefield and Tolmie(2016)}]{rouncefield2016ethnomethodology}
Rouncefield, M.; and Tolmie, P. 2016.
\newblock \emph{Ethnomethodology at work}.
\newblock Routledge.

\bibitem[{Rubin(2022)}]{rubin2022content}
Rubin, V.~L. 2022.
\newblock Content verification for social media: From deception detection to
  automated fact-checking.
\newblock \emph{The SAGE handbook of social media research methods}.

\bibitem[{Si et~al.(2021)Si, Zhou, Li, Shi, and He}]{si2021topic}
Si, J.; Zhou, D.; Li, T.; Shi, X.; and He, Y. 2021.
\newblock Topic-Aware Evidence Reasoning and Stance-Aware Aggregation for Fact
  Verification.
\newblock In \emph{Proceedings of the 59th Annual Meeting of the Association
  for Computational Linguistics and the 11th International Joint Conference on
  Natural Language Processing (Volume 1: Long Papers)}, 1612--1622.

\bibitem[{Slota(2020)}]{slota2020designing}
Slota, S.~C. 2020.
\newblock Designing Across Distributed Agency: Values, participatory design and
  building socially responsible AI.
\newblock \emph{Good Systems-Published Research}.

\bibitem[{Tolmie et~al.(2017)Tolmie, Procter, Randall, Rouncefield, Burger,
  Wong Sak~Hoi, Zubiaga, and Liakata}]{Tolmie2017supporting}
Tolmie, P.; Procter, R.; Randall, D.~W.; Rouncefield, M.; Burger, C.; Wong
  Sak~Hoi, G.; Zubiaga, A.; and Liakata, M. 2017.
\newblock Supporting the use of user generated content in journalistic
  practice.
\newblock In \emph{Proceedings of the 2017 ACM CHI conference on Human Factors
  in Computing Systems}, 3632--3644.

\bibitem[{Wolf(2020)}]{wolf2020ai}
Wolf, C.~T. 2020.
\newblock AI models and their worlds: Investigating data-driven, AI/ML
  ecosystems through a work practices lens.
\newblock In \emph{Sustainable Digital Communities: 15th International
  Conference, iConference 2020, Boras, Sweden, March 23--26, 2020, Proceedings
  15}, 651--664. Springer.

\bibitem[{Zeng, Abumansour, and Zubiaga(2021)}]{zeng2021automated}
Zeng, X.; Abumansour, A.~S.; and Zubiaga, A. 2021.
\newblock Automated fact-checking: A survey.
\newblock \emph{Language and Linguistics Compass}, 15(10): e12438.

\bibitem[{Zhang, Gui, and He(2021)}]{zhang2021supervised}
Zhang, W.; Gui, L.; and He, Y. 2021.
\newblock Supervised contrastive learning for multimodal unreliable news
  detection in covid-19 pandemic.
\newblock In \emph{Proceedings of the 30th ACM International Conference on
  Information \& Knowledge Management}, 3637--3641.

\bibitem[{Zhang, He, and Procter(2023)}]{wenjia2023}
Zhang, W.; He, Y.; and Procter, R. 2023.
\newblock NewsQuote: A Dataset Build on Quote Extraction and Attribution for
  Expert Recommendation.
\newblock In \emph{Proceedings of the International AAAI Conference on Web and
  Social Media}. AAAI Press.

\bibitem[{Zhao et~al.(2023)Zhao, Arana-Catania, Zhu, Kochkina, Gui, Zubiaga,
  Procter, Liakata, and He}]{zhao2023panacea}
Zhao, R.; Arana-Catania, M.; Zhu, L.; Kochkina, E.; Gui, L.; Zubiaga, A.;
  Procter, R.; Liakata, M.; and He, Y. 2023.
\newblock PANACEA: An Automated Misinformation Detection System on COVID-19.
\newblock \emph{arXiv preprint arXiv:2303.01241}.

\bibitem[{Zhu et~al.(2022)Zhu, Fang, Pergola, Procter, and
  He}]{zhu2022disentangled}
Zhu, L.; Fang, Z.; Pergola, G.; Procter, R.; and He, Y. 2022.
\newblock Disentangled Learning of Stance and Aspect Topics for Vaccine
  Attitude Detection in Social Media.
\newblock In \emph{Proceedings of the 2022 Conference of the North American
  Chapter of the Association for Computational Linguistics: Human Language
  Technologies}, 1566--1580.

\end{thebibliography}

\section{Appendix: Semi-structured interview questions}
\noindent\textbf{About you and your team/organisation}
\begin{itemize}
\item What is your background?
\item What is the objective of your team/organisation?
\item What is your role and how long have you been in it?
\end{itemize}

\noindent\textbf{Fact-checking work}
\begin{itemize}
\item What are the main stages in the fact-checking process?
\item How is fact-checking work organised:
\begin{itemize}
\item How are claims discovered?
\item How are decisions made on whether a claim is check-worthy?
\item What kinds of evidence are the most important for a fact-check?
\item What kinds of tools are currently used, for example, to find the evidence?
\item What measures are taken to ensure the process is reliable and timely?
\item What are the main challenges in fact-checking work and why?
\end{itemize}
\item How is the outcome of the fact-checking process used?
\item Do the team members have distinct roles?
\item Do team members undertake any training?
\end{itemize}


\noindent\textbf{New tools to support fact-checking}
\begin{itemize}
\item What are the limitations of the tools that you currently use?
\item At what stage(s) in the fact-checking process would new tools be most valuable?
\item What roles could new tools play?
\item What are the challenges in making them effective and trustworthy?
\item In your opinion is it possible and/or desirable to automate fact-checking?
\end{itemize}
\end{document}